\documentclass[pdflatex,sn-mathphys]{sn-jnl}

\jyear{2022}%

\theoremstyle{thmstyleone}%
\theoremstyle{thmstyletwo}%

\theoremstyle{thmstylethree}%

\begin{document}

\title[Temperature continuously controls the stability of clay slopes]{Temperature continuously controls the stability of clay slopes}


\author[1]{\fnm{Marco} \sur{Loche}}\email{marco.loche@natur.cuni.cz}

\author*[1]{\fnm{Gianvito} \sur{Scaringi}}\email{gianvito.scaringi@natur.cuni.cz}

\affil[1]{\orgdiv{Institute of Hydrogeology, Engineering Geology and Applied Geophysics}, \orgname{Charles University}, \orgaddress{\street{Albertov 6}, \city{Prague}, \postcode{128 43}, \country{Czech Republic}}}



\abstract{With climate change, we are expecting more frequent extreme weather events in many regions worldwide. These events can trigger disruptive, deadly natural hazards, which catch the attention of the media and raise awareness in citizens and policymakers. Floods, wildfires, landslides are the object of a great deal of research. Yet, they remain difficult to predict and handle. Climate change also means warmer temperatures, especially on land. Glaciers melt, permafrost thaws. Everywhere, the ground gets warmer down to increasing depths. At first sight, not a big deal. What can a few extra degrees do? Microbes and fungi becomes more active, chemical equilibria shift. Silent changes are left unnoticed, even by scientists. In landslide studies, the stability of slopes is a balance of weights, water pressures, and mechanical strengths. Above freezing, temperature is left out, yet it should not be. The strength of clays, which frequently are abundant in soils, is sensitive to temperature. With a simple numerical exercise, we show to what extent temperature can condition and even undermine the stability of clay-bearing slopes, across the seasons and under global warming. A silent change that could make a non-extreme weather event have much more severe consequences.
}




\maketitle
\section{Introduction}\label{sec1}

Climate change is disrupting the lives of many, in many forms. It has become a source of concern, and even anxiety for some \cite{wake2022understanding}. A continuous flow of information on natural disasters is saturating broadcasts, magazines, and journals. We are struggling to navigate this global problematic, with no safe ports in sight. With much needed political action, scientists carry the duty of effectively communicating the knowledge and advising on pathways for risk reduction and management \cite{tompkins2004does}. Understanding the physical processes and how they are being altered is crucial to predicting and mitigating the consequences of climate change, or at least slowing down its major effects to give us extra time to adapt.

Extreme weather events are one of these major effects that people are most aware of. Even so, many remain skeptical as for the causal nexus with climate change, as climate change awareness is, for many, a politics-mediated matter \cite{gartner2021}. On the other hand, research on weather extremes and their consequences -- floods, landslides, wildfires, to name a few in the realm of natural hazards -- is flourishing \cite{coumou2012,crozier2010deciphering}. With a very logical Pareto-like approach, scientists attempt to tackle the few main processes that cause the most immediate threats. 

In landslide science (our focus from now on), geologists and engineers describe the stability of slopes via equations of mechanical equilibrium and hydraulic flow \cite{fredlund1977comparison}. Rearrangement of matter (excavation, building), dynamic action (earthquake, blasting), changes in water content and water pressures (rainfall, snowmelt, flooding) are typical events that modelling approaches can directly account for.

\begin{figure}[h]%
\centering
\includegraphics[width=1\textwidth]{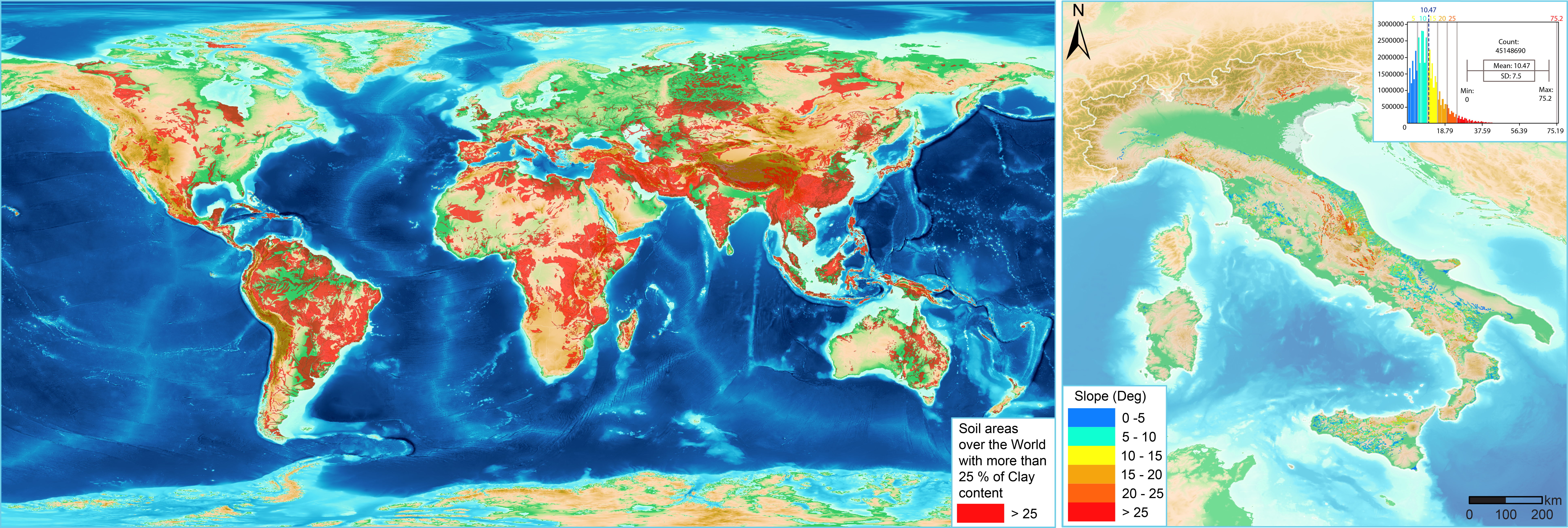}
\caption{Left, Global distribution of soils containing more than 25\% of clay minerals. Right, detail for Italy \cite{ito2017global,shangguan2014global,hengl2017soilgrids250m}}\label{fig1}
\end{figure}

Soils containing clay minerals are the most prone to landsliding owing to their poor mechanical response \cite{skempton1964long}. Key to this response is the low friction between clay particles, combined with their tiny size and platy shape that make them particularly good at interacting with water \cite{mitchell2005fundamentals}. When subject to compression and shearing, which is the case on hillslopes, they easily deform and weaken further, generating smooth weak surfaces of preferential sliding \cite{skempton1985}. Clays and their role in landsliding are, in fact, well studied \cite{leroueil2001,bromhead2013,kasanin2013clay}. Moreover, clays are ubiquitous and abundant all over the globe \cite{ito2017global,shangguan2014global} (Figure \ref{fig1}). 

Clay soils have a pronounced thermal sensitivity in that many of their mechanical and hydraulic properties depend on temperature \cite{scaringi2022thermo}. Besides the obvious effects of phase changes of water on soil volume, stiffness, and permeability, this sensitivity is expressed throughout the whole range of temperatures for liquid water. This is not mainstream knowledge, not even among scientists. In the 1960's, research on the so-called thermo-hydro-mechanical coupling in clays was conducted by civil engineers, tasked to study certain issues related to the large temperature oscillations underneath road pavements in hot and arid regions \cite{mitchell2005fundamentals,richards1969temperature,campanella1968influence}. Recommendations were even made on the need of testing clay soils at temperatures matching those in the field and not just at laboratory standard temperature. However, these works did not make an impact in their field, and were rapidly forgotten.

More recently, research on radioactive waste disposal revamped the interest in thermo-hydro-mechanical effects in clays. The role of engineered soil barriers (mainly bentonites - clays that are particularly reactive) is indeed crucial in impeding or retarding the release of contaminants (including radionuclides) into the groundwater. As radioactive waste produces large amounts of heat while decaying, and can do so for thousands of years, soil barriers need to operate at elevated temperatures. At the same time, they need to withstand the large pressure from the surrounding rock (they are installed deep underground) and accommodate for water slowly filling their pores (to enhance their capabilities, the barriers are installed in almost dry condition) \cite{delage2010clays,sun2020water,villar2004influence}. These studies mainly revolve around changes in water sorption capacity and compressibility with temperature. However, depending on boundary conditions (whether deformation is possible, whether water is available), changes in temperature may alter internal stresses and water pressures, triggering flows and deformations \cite{rotta2017thermally}. The internal mechanisms are manifold and mutually interacting, with both positive (reinforcing) and negative (cancelling each other out) feedback. Water and clay minerals display different thermal expansions, which results in changes in water pressures; as temperature increases, interparticle forces change as water becomes less viscous and more electrically conductive, and clay surfaces are less capable of withholding it \cite{mitchell2005fundamentals}. Moreover, heating and cooling cycles in soft soils can produce stiffening and a net shrinkage (counterintuitively) \cite{campanella1968influence, tang2008thermo}, while the opposite holds true in overconsolidated soils. 

The shear strength is not a key parameter in engineered soil barriers as they are not meant to undergo large deformations. Therefore, research on the thermal sensitivity of shear strength parameters has historically been less developed \cite{scaringi2022thermo, loche2022heating, loche2021infrared, loche2022surface}. As a matter of fact, whether the shear strength increases or decreases with temperature in a clay soil depends on both its nature and structure, the latter imposed on the soil by its stress and thermal history \cite{hueckel2009explaining}. The so-called residual shear strength, the sole available in soils after large shear strains (and thus in basal slip surfaces or landslide shear zones), is proper of the material and does not depend on its history. Evaluating the thermal sensitivity of the residual shear strength is, therefore, more straightforward. Even so, the literature is particularly poor in such evaluations because temperature has not generally been deemed a matter of concern in large-deformation problems (lateral strength of foundation piles, basal strength of typical-size landslides) \cite{Reichenbach2018}. In one notable example, the seasonal reactivation of a number of slow-moving landslides in a clay-rich formation in Japan, occurring in late autumn in absence of precipitation or stress changes, was attributed to soil weakening caused by cooling \cite{shibasaki2016experimental}. This was confirmed by systematic experiments exploring the role of clay mineralogy on the thermal response \cite{shibasaki2017temperature,loche2022heating}, which pointed out also the role of the rate of shearing and thus prospected different effects on slow and fast movements \cite{scaringi2022thermo}.

As aforementioned, soil slope stability models only focus on hydro-mechanics. Yet more advanced models exist (developed for other applications, such as for modelling clay barriers) that explicitly account for thermo-hydro-mechanical processes \citep{mavsin2017coupled}. These models, however, have not yet been developed for landslides and typically remain confined to small-scale domains also owing to computational burdens. In absence of a modelling tool and a widespread knowledge on the thermal sensitivity of soils -- but justified by the dampening role of soil heterogeneity and by temperature oscillations fading at large depths -- even in climate change scenarios, expected changes in slope stability (in temperate regions) are mainly those due to changing hydrological inputs \cite{crozier2010deciphering,gariano2016landslides,elia2017numerical}. 

To explore the individual role of temperature in clay slope stability, we performed a virtual experiment on a typical, linear slope profile. We sequentially coupled a simple heat diffusion analysis in a homogeneous, water-saturated medium, with a conventional slope stability analysis with subdomains of different strength generated according to the temperature. For selected slip surfaces, we evaluated changes in the factor of safety resulting from seasonal changes in surface temperature. We then explored how an increase in temperature over decades (simulating climate change) would further affect the stability. Our argument is that, before dismissing thermal effects, a simple numerical analysis combined with few laboratory experiments can tell whether an in-depth investigation and fully-coupled analyses are necessary. In absence of this quantitative check, climate change effects on slope stability (and consequent chains of hazards) may be significantly misestimated, with consequences for risk assessments and mitigation strategies.

\section{Results and Discussion}\label{sec2}

\begin{figure}[htbp]%
\centering
\includegraphics[width=0.6\textwidth]{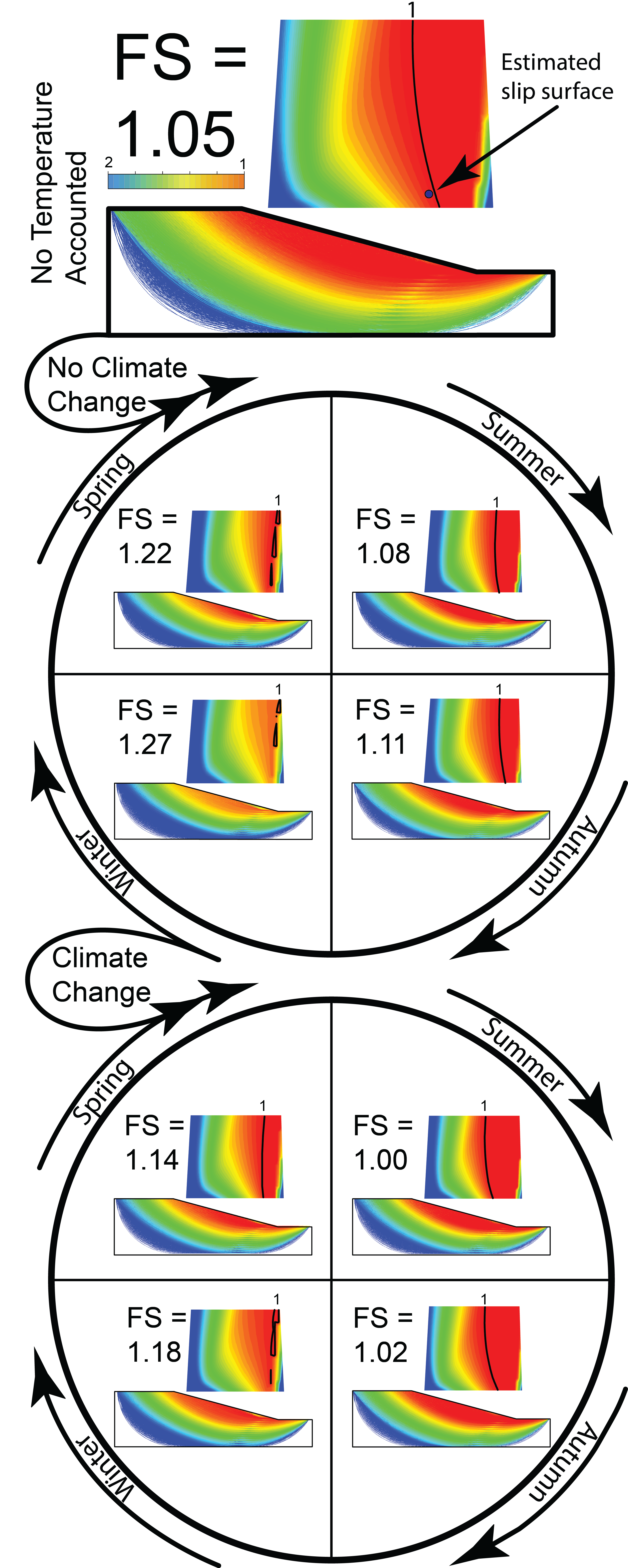}
\caption{Factor of safety for the model slopes in absence of thermal sensitivity (top), with thermal sensitivity (centre, according to the seasons), and with thermal sensitivity and climate change (top, according to the seasons, after a warming of 5°C).}\label{fig2}
\end{figure}

Figure \ref{fig2} summarises our two key results, that is that: 1) not accounting for the temperature-shear strength relationship in slope stability analysis can lead to a predicted factor of safety that differs significantly from the actual one under model conditions; and 2) not accounting for warming due to climate change leads to a significant misestimation of future slope stability. In our model simulations, a change in factor of safety by up to 27\% was evaluated, considering a homogeneous soil slope made of a thermally-sensitive material (an active clay), a sinusoidal thermal forcing representing the alternation of seasons (24°C of amplitude) in a temperate or warm, non-freezing climate, and no other forcings (no changes in water content, pore water pressures, mechanical loads). Our analysis is certainly simplified and, obviously, thermal forcing is always coupled at least with changes in hydraulic conditions. However, we only aimed at displaying the magnitude of the direct effect of temperature on the shear strength. In real conditions, other forcings may diminish but also enhance the role of soil thermal sensitivity.

More in detail, in the top of Figure \ref{fig2}, the result of a classic slope stability analysis is reported. The colour shades represent the factors of safety associated with potential slip surfaces in the slope (whose centres also are displayed in the shaded quadrilateral above it). Deeper slip surfaces are less likely than shallower ones. However, the latter represent possible local and shallow failures, and are not significant in terms of global slope stability. We have not considered them because suction, cohesion, and/or partial saturation would typically prevent them. To explore global failures, but also to make sub-seasonal temperature fluctuations (not considered in the chosen thermal forcing) insignificant, we only considered surfaces reaching a maximum depth of at least 6 m. Nonetheless, the analysis of the effect of temperature is valid for any of the potential slip surfaces; we chose a specific one as a didactic example. As shown in the figure, along this slip surface, a factor of safety of 1.05 is evaluated. The value is a function of the chosen value for the shear strength. Imagining that this is a preexisting slip surface (rather than a randomly chosen one), the residual shear strength parameter would be the one to apply. This does not alter the generality of the result. This strength parameter can be thought as having been evaluated via appropriate laboratory experiments (e.g., direct or ring-shear tests), performed on water-saturated undisturbed or reconstituted soil specimens sampled through boreholes from the slip surface. These experiments would have likely been performed without temperature control, at room temperature (say at 20°C).

The second part of Figure \ref{fig2} links the slope stability analysis with the changing temperature in the slope (the pattern of temperature with depth is displayed in the methods). We set our analysis at an imaginary location where the average annual temperature is 15°C. This 5°C difference between the laboratory and field conditions can already cause a difference in soil strength by up to 7.5\%. If the material exhibits a "negative temperature effect" (i.e., it is stronger at lower temperatures), as in the case shown in the figure, the slope will actually be more stable than estimated, as long as field temperatures are lower than those in the laboratory. However, the opposite will be true if the material shows a "positive temperature effect" (i.e., it gets stronger at higher temperatures). Note that a discrepancy by 7.5\% between model estimates and actual field conditions may be acceptable in simplified analyses owing to other sources of uncertainties and consequent safety margins in design (e.g., in the construction of retaining walls). On the other hand, a source of uncertainty unaccounted for in models make them less reliable, and uncertainties can propagate further and amplify (e.g., in risk assessment or zoning). Obviously, these considerations should be done case-by-case, knowing field temperatures and materials, to judge whether soil thermal sensitivity can be neglected. More in detail, the figure also shows how the factor of safety calculated for the selected slip surface changes during the seasons. For the chosen thermal response of soil, the lowest safety factor is in summer and the highest in winter, with spring and autumn conditions being asymmetrical despite having assigned the same value of surface temperature. This is due to the time required for the propagation of heat in the underground. In other words, the safety factor in autumn is still close to the one in summer because the ground keeps the thermal energy accumulated during the summer. Once again, this picture would be different if the soil had a thermal sensitivity of opposite sign. In reality, actual factors of safety also depend on the changing hydraulic conditions, thus not only the average temperature but also how it changes during the seasons and how this couples with the patterns of precipitation needs to be considered. In fact, a soil that is weaker in summer would suffer particularly from summer storms capable of quickly saturating and/or raising pore water pressures, but would be stronger in winter and capable to better withstand wintertime rain or snow. The opposite holds true for a soil that is weaker in winter. The overall behaviour and pattern of landslides may be influenced by this interaction between soil type/response and precipitations that can be either distributed or concentrated in a specific season.

Finally, the bottom part of Figure \ref{fig2} shows what could happen in terms of slope stability during the seasons if a warming by 5°C occurs. The result, easy to guess, is that because of warming, progressively propagating from the surface to the underground, the soil weakens and lower factors of safety are evaluated. In the specific case, failure (or reactivation for an existing surface) would occur along the observed slip surface in summertime, without the need of any additional (hydro-mechanical) forcing, or in autumn perhaps because of a mild rainfall event. Consider that a warming by 5°C in the coming decades is not unrealistic and does not even correspond to a worst-case scenario; certain temperate regions have already witnessed a warming by 2°C in past decades, and the trend is not slowing down. Obviously, once again, the direct effect of temperature on strength needs to be coupled with expected changes in hydro-meteorological patterns. The imaginary location chosen for this virtual experiment may witness less total precipitation (which is good for stability) but also more extreme events, such as summer storms. Therefore, judging whether there will be more or less landslides in soil slopes simply by performing temperature-sensitive slope stability analyses is not possible, but our model results clearly point out that temperature could matter and could even make a difference between a "correct" and a "wrong" (false negative or false positive) prediction of stability/instability.

\begin{figure}[h]%
\centering
\includegraphics[width=0.7\textwidth]{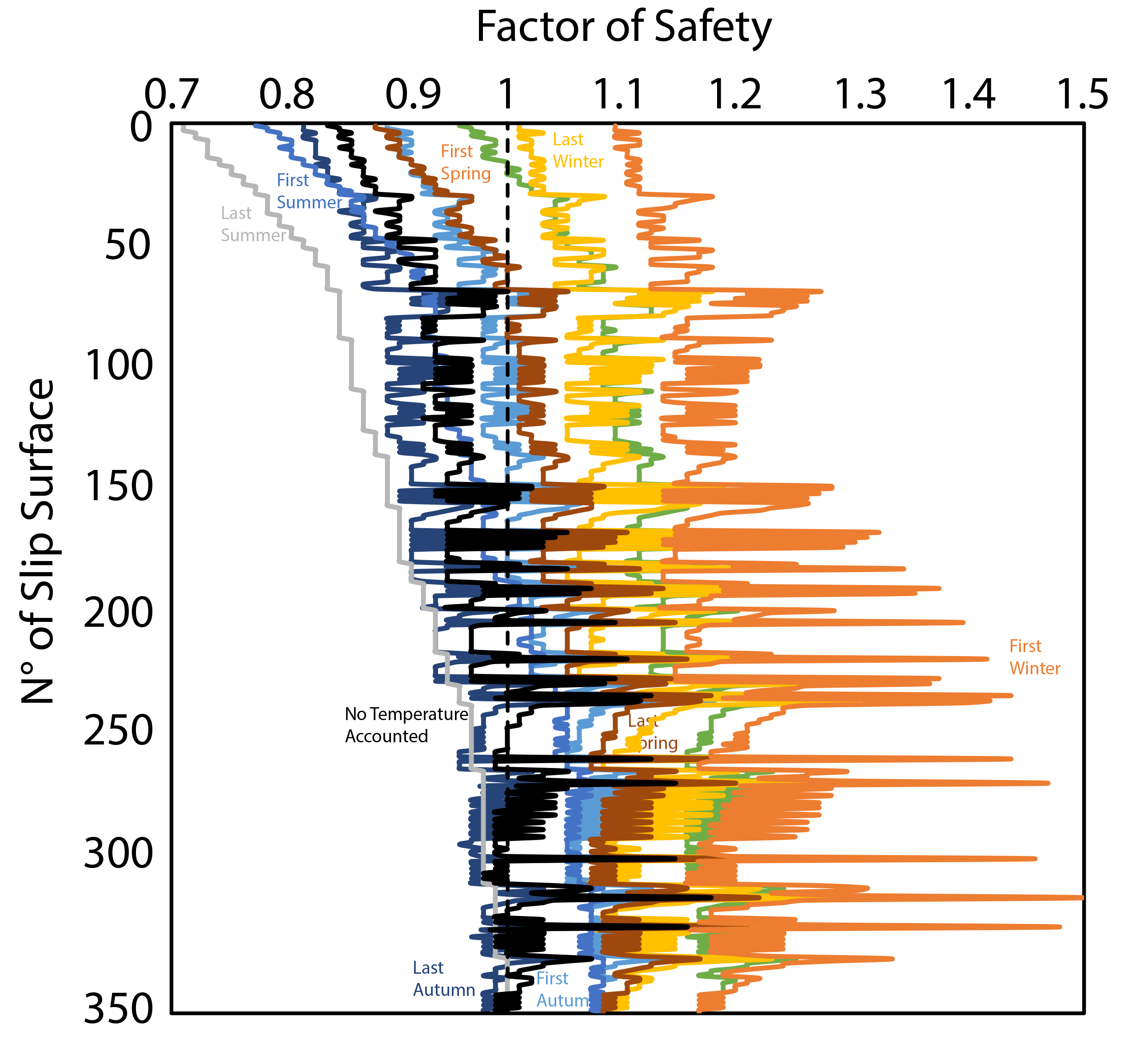}
\caption{Factor of safety along potential slip surfaces (restricted to surfaces with a maximum depth of at least 6 m and values from 0.7 to 1.5) in absence of thermal sensitivity, according to the seasons (labelled as first spring, etc.), and after a warming of 5°C (labelled as last spring, etc.).}\label{fig3}
\end{figure}

For completeness, results relative to many potential slip surfaces are summarised in Figure \ref{fig3}). The figure essentially shows that, for 350 surfaces analysed with factors of safety close to one, accounting for thermal sensitivity and then warming due to climate change can significantly change the predicted stability condition. Almost half of the surfaces, stable in absence of warming or without accounting for thermal sensitivity, are deemed unstable if the latter is considered. Note that, the closer a surface is to failure, the smaller, further perturbation will be necessary to trigger instability. This "mild" perturbation could be represented by a rainfall event with a short return period.

Once again, it should be stressed that we presented a simple numerical exercise, but we used knowledge derived from laboratory experiments that had not been included in slope stability or landslide modelling thus far. The choice of model geometry was generic: the model output basically only depends on the choice of the slope angle in relation to the soil's friction angle. Even in absence of predicted failure, a change in factor of safety of the same proportion as that evaluated above would be displayed, which is solely depending on the thermal sensitivity of the soil. We used the largest value of sensitivity (a change in strength by 1.5\% per °C) according to experimental results for a smectitic clay to depict a "worst-case scenario" in terms of "missed" temperature effects. We stressed, however, that soil behaviour in shear is usually dominated by the response of the clay component at usual landslide stress levels, so that even a relatively small proportion of clay in soil can produce shear behaviours proper of a pure clay. For a failure surface that can be considered not so shallow ($>$ 6 m), changes in temperature in the ground, as a consequences of seasons and warming in a temperate climate (i.e., in absence of freezing), can cause important changes in factor of safety as a direct result of the soil thermal sensitivity. Notably, even though at a depth of 6 m seasonal temperature oscillations are quite dampened, a large portion of the slip surface lies at shallow depths, where the amplitude of oscillations is larger. In field conditions, where the slip surface is not confined to a longitudinal plane as in our two-dimensional modelling, this is even more true. Furthermore, preferential air and water flow pathways such as discontinuities would enhance the amplitude of temperature oscillations (while deeper or adjacent aquifers would suppress them owing to their large heat capacity). Soil spatial heterogeneity would reduce the overall thermal sensitivity and should be analyzed case-by-case. The effect of precipitation on soil hydro-mechanics as well as temperature distribution also should be evaluated. However, it remains clear that temperature (as an independent variable conditioning slope stability) can play a significant role, both alone and in combination with other forcings. We argue, therefore, that temperature should be considered systematically in the analysis of the stability of slopes in clay-bearing soils in temperate regions. Evaluating the soil's thermal sensitivity and considering the differences between laboratory and field conditions should be the first step, and at least sequentially coupled simulations of temperature and slope stability should be performed in case of large thermal sensitivity. The third step, in case of important effects on stability, would be to carry out fully-coupled thermo-hydro-mechanical analyses, for which suitable modelling approaches are not yet available. At a broader scale, neglecting the direct effect of temperature on soil strength and hence slope stability may have consequences in risk assessments and lead to insufficiently informed strategies for risk mitigation in the context of climate change.

\section{Methods}\label{sec3}

Temperature in the shallow subsurface (down to depths of 10--20 m) is essentially the result of a dynamic energy balance between the heat coming from the inner Earth and that penetrating from the surface via irradiation, conduction, and convection. The actual dynamics are rather complex and depend on local conditions in the crust as well as the specific location and climate (latitude, solar irradiation, air temperature and humidity, wind, slope orientation and surface morphology, vegetation). Thermal conductivity and heat capacity depend on soil composition and moisture content. Convection in the fluid phases depends on porosity and pore network structure, and degree of saturation. Groundwater flow can facilitate heat transport in its direction, increasing or decreasing the depth of penetration of temperature fluctuations. Once again, local specificities warrant case-by-case analyses.

To provide a sample analysis that quantifies up to what extent temperature alone can control the stability of a slope, we worked with a simple geometry consisting of a 1:4 slope and an underground domain up to 20 m deep consisting of a homogeneous, water-saturated material (with a groundwater table corresponding to the ground surface) with no cohesion and a friction angle set at 27° at the temperature $T_{0}$ = 20°C. We used a simple formulation to describe the thermal sensitivity:

\begin{equation}
{\varphi}(T)={\varphi}_{T_0} \alpha (T-T_0).\label{eq1}
\end{equation}

where

\begin{equation}
\left[\frac{\Delta \varphi}{\Delta T}\right] \leq {0}\label{eq2}
\end{equation}
\\
    
that is we chose $\alpha < 0$ and, in particular, $\alpha < 0.015$ consistently with the largest thermal sensitivity evaluated in the literature in the range of positive temperatures up to ~50°C \cite{shibasaki2016experimental,shibasaki2017temperature,loche2022heating}. Note that the sign of $\alpha$ depends on soil mineralogy, with the largest effects evaluated in clays. In particular, $\alpha < 0$, that is a cooling-induced weakening, was evaluated in smectitic clays during slow shearing ($v < ~1$ mm/min) and in non-smectitic clays during fast shearing ($v > ~1$ mm/min). Conversely, $\alpha > 0$, that is a heating-induced weakening, was evaluated in smectitic clays during fast shearing and in non-smectitic clays during slow shearing. Note that we used the largest possible value for $\alpha$, that should be attributable to pure clays. However, in clay-non-clay mixtures, the available shearing resistance is controlled by the component that occupies the largest volume, and hence the largest part of the shear zone. Under typical landslide stress levels (0--150 kPa), the specific volume of the least active clay minerals can be significantly larger than that of non-clay minerals (e.g., quartz sand). For the most active clays (e.g., montmorillonite), the specific volume can even be an order of magnitude larger. As such, even a small percentage in weight of an active clay (e.g., 10--20\%) can be sufficient to completely control the mixture's mechanical behaviour \cite{skempton1985}, including its shear strength and, consequently, the thermal sensitivity of the latter. 

As for the specific value of friction angle chosen, it does not belong to a specific material. It only has to be considered in relation to the chosen slope geometry: our choice was such to produce safety factors close to unity, and hence explore changes in slope stability in not-definitely-stable slopes. While this may seem restrictive, it should be considered that the abundance of landslides in clay soils suggests that such conditions are frequent. Moreover, erosional processes naturally tend to produce landscapes with just enough margin of safety against global failure. We stress that our analyses could have been conducted with any slope angle -- friction angle couple.

In terms of boundary conditions, we fixed the temperature at the bottom boundary and let the temperature at the surface fluctuate according to a sinusoidal forcing representing the alternation of seasons, assuming an annual average temperature of 15°C and a semi-amplitude of 12°C (i.e., an average temperature of 27°C on the hottest day and 3°C on the coldest). Clearly, the sinusoidal forcing is an approximation, but it should be noted that short-period oscillations or perturbations hardly penetrate past the first metres of ground. As for the average and amplitude values, these may be realistic for a temperate/warm climate, but should be specialised according to the specific location. 

\begin{figure}[h]%
\centering
\includegraphics[width=0.7\textwidth]{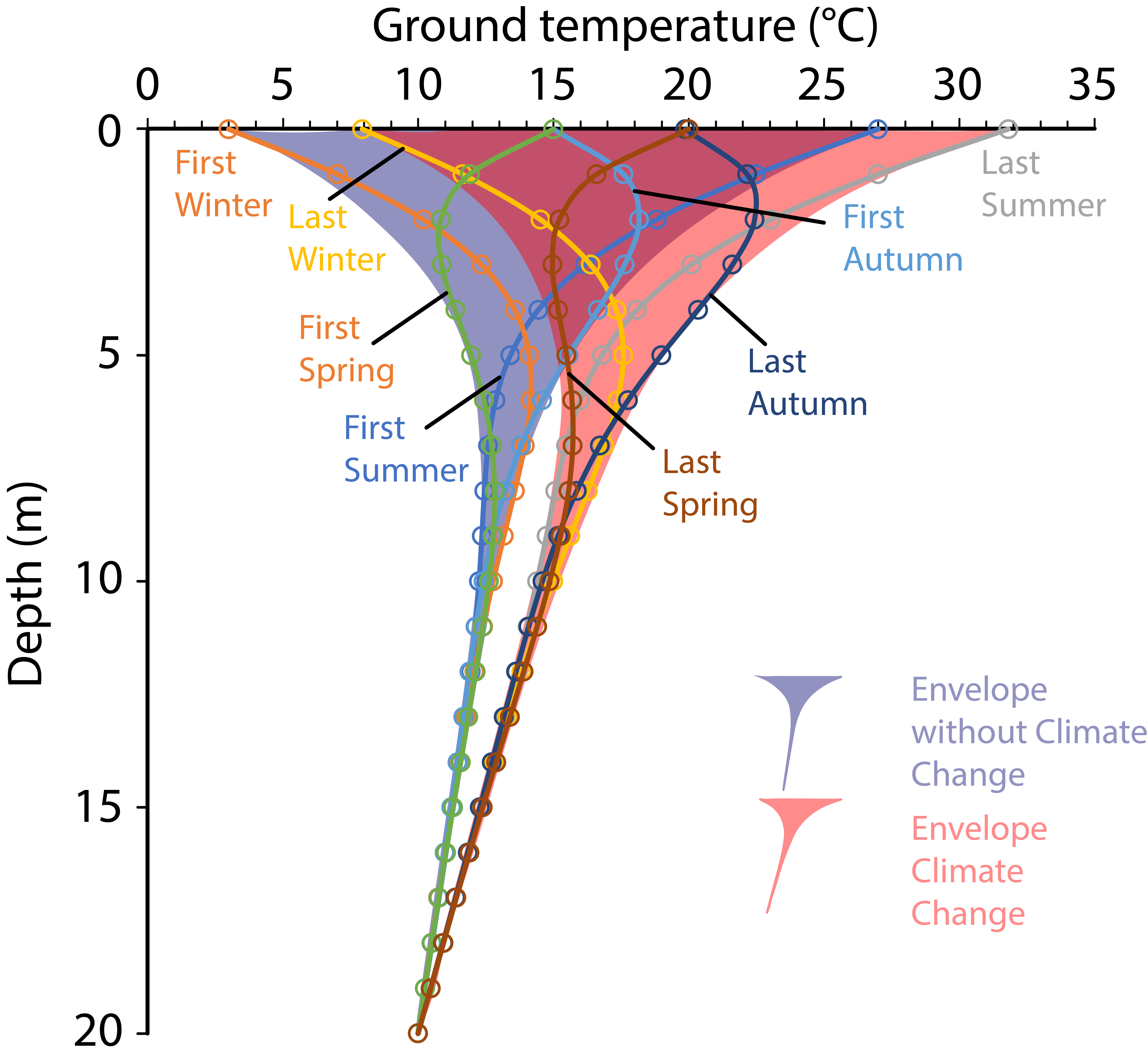}
\caption{Depth profiles of ground temperature. Seasonal variations in ground temperature in the baseline case ("no climate change") and after a warming of 5°C ("climate change"}\label{fig4}
\end{figure}

We used the software package GeoStudio distributed by Seequent, which is a widely used tool in geotechnical engineering, and in particular the TEMP/W, SEEP/W, and SLOPE/W subpackages. We ran a baseline analysis in which the effect of temperature was not considered and a constant value of friction angle was used across the domain. Then, we imposed the sinusoidal forcing and let the domain attain a steady state. We took snapshots of temperature in four seasons, and assigned values of friction angle to subdomains according to the above-described equation. To simulate the warming due to climate change, we finally imposed a gradual increase of average temperature by 5°C (without changing the amplitude of the forcing) over a sufficiently long period (30 years) to allow for the entire domain to reach a new steady state. Longer periods would not alter the final result, whereas shorter periods would result in changes in deep temperatures lagging behind (by up to 5 years in our model setup) those at the surface. Typical ground temperature profiles resulting from the TEMP/W analyses are displayed in Figure \ref{fig4}. These profiles match with experimentally measured ones \cite{shibasaki2016experimental,kurylyk2015}.

Slope stability was evaluated in terms of a safety factor (the ratio between resisting forces and forces driving instability) using the Bishop method of slices, but other methods available in the software would all give comparable results. We limited the analysis to potential slip surfaces, of circular shape as a consequence of the chosen method, having a depth of at least 6 m in their deepest point. This choice was made to automatically discard very shallow surfaces (clearly affected by temperature oscillations but also by important changes in moisture and pore water pressures - neglected in this work), and surfaces delineating only local failures. We were, in fact, interested in global failure surfaces (affecting most part of the slope) and reaching depths such that portions of ground at significantly different temperatures would be involved.

\backmatter
\bmhead{Acknowledgments}

This work was financially supported by the Grant Agency of the Czech Republic (GA{\v C}R; Grant No. 20-28853Y), the Fund for international mobility of researchers at Charles University (MSCA-IF IV; Project No. CZ.02.2.69/0.0/0.0/20\_079/0017987), and the Charles University Grant Agency (GAUK; Project No. 337121).

\section*{Declarations}

\begin{itemize}
\item Funding: 
See acknowledgment.
\item Conflict of interest/Competing interests:
The authors declare that there is no conflict of interest.
\item Availability of data and materials:
This work was built upon freely available datasets and tools.
\item Code availability: 
Not applicable 
\item Authors' contributions:
GS supervised the research; GS, ML developed the research idea; ML obtained the input data; ML elaborated the data and implemented the model with help from GS; ML wrote the initial draft; both authors revised and approved the manuscript.

\end{itemize}

\begin{appendices}

\section{}\label{secA1}




\end{appendices}


\bibliography{sn-bibliography}


\begin{thebibliography}{34}
\ifx \bisbn   \undefined \def \bisbn  #1{ISBN #1}\fi
\ifx \binits  \undefined \def \binits#1{#1}\fi
\ifx \bauthor  \undefined \def \bauthor#1{#1}\fi
\ifx \batitle  \undefined \def \batitle#1{#1}\fi
\ifx \bjtitle  \undefined \def \bjtitle#1{#1}\fi
\ifx \bvolume  \undefined \def \bvolume#1{\textbf{#1}}\fi
\ifx \byear  \undefined \def \byear#1{#1}\fi
\ifx \bissue  \undefined \def \bissue#1{#1}\fi
\ifx \bfpage  \undefined \def \bfpage#1{#1}\fi
\ifx \blpage  \undefined \def \blpage #1{#1}\fi
\ifx \burl  \undefined \def \burl#1{\textsf{#1}}\fi
\ifx \doiurl  \undefined \def \doiurl#1{\url{https://doi.org/#1}}\fi
\ifx \betal  \undefined \def \betal{\textit{et al.}}\fi
\ifx \binstitute  \undefined \def \binstitute#1{#1}\fi
\ifx \binstitutionaled  \undefined \def \binstitutionaled#1{#1}\fi
\ifx \bctitle  \undefined \def \bctitle#1{#1}\fi
\ifx \beditor  \undefined \def \beditor#1{#1}\fi
\ifx \bpublisher  \undefined \def \bpublisher#1{#1}\fi
\ifx \bbtitle  \undefined \def \bbtitle#1{#1}\fi
\ifx \bedition  \undefined \def \bedition#1{#1}\fi
\ifx \bseriesno  \undefined \def \bseriesno#1{#1}\fi
\ifx \blocation  \undefined \def \blocation#1{#1}\fi
\ifx \bsertitle  \undefined \def \bsertitle#1{#1}\fi
\ifx \bsnm \undefined \def \bsnm#1{#1}\fi
\ifx \bsuffix \undefined \def \bsuffix#1{#1}\fi
\ifx \bparticle \undefined \def \bparticle#1{#1}\fi
\ifx \barticle \undefined \def \barticle#1{#1}\fi
\bibcommenthead
\ifx \bconfdate \undefined \def \bconfdate #1{#1}\fi
\ifx \botherref \undefined \def \botherref #1{#1}\fi
\ifx \url \undefined \def \url#1{\textsf{#1}}\fi
\ifx \bchapter \undefined \def \bchapter#1{#1}\fi
\ifx \bbook \undefined \def \bbook#1{#1}\fi
\ifx \bcomment \undefined \def \bcomment#1{#1}\fi
\ifx \oauthor \undefined \def \oauthor#1{#1}\fi
\ifx \citeauthoryear \undefined \def \citeauthoryear#1{#1}\fi
\ifx \endbibitem  \undefined \def \endbibitem {}\fi
\ifx \bconflocation  \undefined \def \bconflocation#1{#1}\fi
\ifx \arxivurl  \undefined \def \arxivurl#1{\textsf{#1}}\fi
\csname PreBibitemsHook\endcsname

\bibitem{wake2022understanding}
\begin{botherref}
\oauthor{\bsnm{Wake}, \binits{B.}}:
Understanding eco-anxiety.
Nature Climate Change,
1--1
(2022)
\end{botherref}
\endbibitem

\bibitem{tompkins2004does}
\begin{botherref}
\oauthor{\bsnm{Tompkins}, \binits{E.L.}},
\oauthor{\bsnm{Adger}, \binits{W.N.}}:
Does adaptive management of natural resources enhance resilience to climate
  change?
Ecology and society
\textbf{9}(2)
(2004)
\end{botherref}
\endbibitem

\bibitem{gartner2021}
\begin{botherref}
\oauthor{\bsnm{Gartner}, \binits{L.}},
\oauthor{\bsnm{Schoen}, \binits{H.}}:
Experiencing climate change: revisiting the role of local weather in affecting
  climate change awareness and related policy preferences.
Climatic Change
\textbf{167}(31)
(2021).
\doiurl{10.1007/s10584-021-03176-z}
\end{botherref}
\endbibitem

\bibitem{coumou2012}
\begin{barticle}
\bauthor{\bsnm{Coumou}, \binits{D.}},
\bauthor{\bsnm{Rahmstorf}, \binits{S.}}:
\batitle{A decade of weather extremes}.
\bjtitle{Nature Climate Change}
\bvolume{2},
\bfpage{491}--\blpage{496}
(\byear{2012})
\end{barticle}
\endbibitem

\bibitem{crozier2010deciphering}
\begin{barticle}
\bauthor{\bsnm{Crozier}, \binits{M.J.}}:
\batitle{Deciphering the effect of climate change on landslide activity: A
  review}.
\bjtitle{Geomorphology}
\bvolume{124}(\bissue{3-4}),
\bfpage{260}--\blpage{267}
(\byear{2010})
\end{barticle}
\endbibitem

\bibitem{fredlund1977comparison}
\begin{barticle}
\bauthor{\bsnm{Fredlund}, \binits{D.G.}},
\bauthor{\bsnm{Krahn}, \binits{J.}}:
\batitle{Comparison of slope stability methods of analysis}.
\bjtitle{Canadian geotechnical journal}
\bvolume{14}(\bissue{3}),
\bfpage{429}--\blpage{439}
(\byear{1977})
\end{barticle}
\endbibitem

\bibitem{skempton1964long}
\begin{barticle}
\bauthor{\bsnm{Skempton}, \binits{A.W.}}:
\batitle{Long-term stability of clay slopes}.
\bjtitle{Geotechnique}
\bvolume{14}(\bissue{2}),
\bfpage{77}--\blpage{102}
(\byear{1964})
\end{barticle}
\endbibitem

\bibitem{mitchell2005fundamentals}
\begin{bbook}
\bauthor{\bsnm{Mitchell}, \binits{J.K.}},
\bauthor{\bsnm{Soga}, \binits{K.}}:
\bbtitle{Fundamentals of Soil Behavior}.
\bpublisher{John Wiley \& Sons New York}, \blocation{???}
(\byear{2005})
\end{bbook}
\endbibitem

\bibitem{skempton1985}
\begin{barticle}
\bauthor{\bsnm{Skempton}, \binits{A.W.}}:
\batitle{Residual strength of clays in landslides, folded strata and the
  laboratory}.
\bjtitle{Geotechnique}
\bvolume{35},
\bfpage{3}--\blpage{18}
(\byear{1985})
\end{barticle}
\endbibitem

\bibitem{leroueil2001}
\begin{barticle}
\bauthor{\bsnm{Leroueil}, \binits{S.}}:
\batitle{Natural slopes and cuts: movement and failure mechanisms}.
\bjtitle{Geotechnique}
\bvolume{51},
\bfpage{197}--\blpage{243}
(\byear{2001})
\end{barticle}
\endbibitem

\bibitem{bromhead2013}
\begin{barticle}
\bauthor{\bsnm{Bromhead}, \binits{E.N.}}:
\batitle{Reflections on the residual strength of clay soils, with special
  reference to bedding-controlled landslides}.
\bjtitle{Quarterly Journal of Engineering Geology and Hydrogeology}
\bvolume{46},
\bfpage{132}--\blpage{155}
(\byear{2013})
\end{barticle}
\endbibitem

\bibitem{kasanin2013clay}
\begin{barticle}
\bauthor{\bsnm{Kasanin-Grubin}, \binits{M.}}:
\batitle{Clay mineralogy as a crucial factor in badland hillslope processes}.
\bjtitle{Catena}
\bvolume{106},
\bfpage{54}--\blpage{67}
(\byear{2013})
\end{barticle}
\endbibitem

\bibitem{ito2017global}
\begin{barticle}
\bauthor{\bsnm{Ito}, \binits{A.}},
\bauthor{\bsnm{Wagai}, \binits{R.}}:
\batitle{Global distribution of clay-size minerals on land surface for
  biogeochemical and climatological studies}.
\bjtitle{Scientific data}
\bvolume{4}(\bissue{1}),
\bfpage{1}--\blpage{11}
(\byear{2017})
\end{barticle}
\endbibitem

\bibitem{shangguan2014global}
\begin{barticle}
\bauthor{\bsnm{Shangguan}, \binits{W.}},
\bauthor{\bsnm{Dai}, \binits{Y.}},
\bauthor{\bsnm{Duan}, \binits{Q.}},
\bauthor{\bsnm{Liu}, \binits{B.}},
\bauthor{\bsnm{Yuan}, \binits{H.}}:
\batitle{A global soil data set for earth system modeling}.
\bjtitle{Journal of Advances in Modeling Earth Systems}
\bvolume{6}(\bissue{1}),
\bfpage{249}--\blpage{263}
(\byear{2014})
\end{barticle}
\endbibitem

\bibitem{scaringi2022thermo}
\begin{botherref}
\oauthor{\bsnm{Scaringi}, \binits{G.}},
\oauthor{\bsnm{Loche}, \binits{M.}}:
A thermo-hydro-mechanical approach to soil slope stability under climate
  change.
Geomorphology,
108108
(2022)
\end{botherref}
\endbibitem

\bibitem{richards1969temperature}
\begin{botherref}
\oauthor{\bsnm{Richards}, \binits{F.}}:
Temperature effects on the engineering properties and behavior of soils.
Special Report
(103),
9
(1969)
\end{botherref}
\endbibitem

\bibitem{campanella1968influence}
\begin{barticle}
\bauthor{\bsnm{Campanella}, \binits{R.G.}},
\bauthor{\bsnm{Mitchell}, \binits{J.K.}}:
\batitle{Influence of temperature variations on soil behavior}.
\bjtitle{Journal of the Soil Mechanics and Foundations Division}
\bvolume{94}(\bissue{3}),
\bfpage{709}--\blpage{734}
(\byear{1968})
\end{barticle}
\endbibitem

\bibitem{delage2010clays}
\begin{barticle}
\bauthor{\bsnm{Delage}, \binits{P.}},
\bauthor{\bsnm{Cui}, \binits{Y.-J.}},
\bauthor{\bsnm{Tang}, \binits{A.M.}}:
\batitle{Clays in radioactive waste disposal}.
\bjtitle{Journal of Rock Mechanics and Geotechnical Engineering}
\bvolume{2}(\bissue{2}),
\bfpage{111}--\blpage{123}
(\byear{2010})
\end{barticle}
\endbibitem

\bibitem{sun2020water}
\begin{barticle}
\bauthor{\bsnm{Sun}, \binits{H.}},
\bauthor{\bsnm{Ma{\v{s}}{\'\i}n}, \binits{D.}},
\bauthor{\bsnm{Najser}, \binits{J.}},
\bauthor{\bsnm{Scaringi}, \binits{G.}}:
\batitle{Water retention of a bentonite for deep geological radioactive waste
  repositories: High-temperature experiments and thermodynamic modeling}.
\bjtitle{Engineering Geology}
\bvolume{269},
\bfpage{105549}
(\byear{2020})
\end{barticle}
\endbibitem

\bibitem{villar2004influence}
\begin{barticle}
\bauthor{\bsnm{Villar}, \binits{M.V.}},
\bauthor{\bsnm{Lloret}, \binits{A.}}:
\batitle{Influence of temperature on the hydro-mechanical behaviour of a
  compacted bentonite}.
\bjtitle{Applied clay science}
\bvolume{26}(\bissue{1-4}),
\bfpage{337}--\blpage{350}
(\byear{2004})
\end{barticle}
\endbibitem

\bibitem{rotta2017thermally}
\begin{barticle}
\bauthor{\bsnm{Rotta~Loria}, \binits{A.F.}},
\bauthor{\bsnm{Laloui}, \binits{L.}}:
\batitle{Thermally induced group effects among energy piles}.
\bjtitle{G{\'e}otechnique}
\bvolume{67}(\bissue{5}),
\bfpage{374}--\blpage{393}
(\byear{2017})
\end{barticle}
\endbibitem

\bibitem{tang2008thermo}
\begin{barticle}
\bauthor{\bsnm{Tang}, \binits{A.-M.}},
\bauthor{\bsnm{Cui}, \binits{Y.-J.}},
\bauthor{\bsnm{Barnel}, \binits{N.}}:
\batitle{Thermo-mechanical behaviour of a compacted swelling clay}.
\bjtitle{G{\'e}otechnique}
\bvolume{58}(\bissue{1}),
\bfpage{45}--\blpage{54}
(\byear{2008})
\end{barticle}
\endbibitem

\bibitem{loche2022heating}
\begin{botherref}
\oauthor{\bsnm{Loche}, \binits{M.}},
\oauthor{\bsnm{Scaringi}, \binits{G.}}:
Heating-induced strengthening or weakening of clays during slow to fast
  shearing at landslide stress levels.
arXiv preprint arXiv:2211.05058
(2022)
\end{botherref}
\endbibitem

\bibitem{loche2021infrared}
\begin{botherref}
\oauthor{\bsnm{Loche}, \binits{M.}},
\oauthor{\bsnm{Scaringi}, \binits{G.}},
\oauthor{\bsnm{Blahůt}, \binits{J.}},
\oauthor{\bsnm{Melis}, \binits{M.T.}},
\oauthor{\bsnm{Funedda}, \binits{A.}},
\oauthor{\bsnm{Da~Pelo}, \binits{S.}},
\oauthor{\bsnm{Erbì}, \binits{I.}},
\oauthor{\bsnm{Deiana}, \binits{G.}},
\oauthor{\bsnm{Meloni}, \binits{M.A.}},
\oauthor{\bsnm{Cocco}, \binits{F.}}:
An infrared thermography approach to evaluate the strength of a rock cliff.
Remote Sensing
\textbf{13}(7)
(2021).
\doiurl{10.3390/rs13071265}
\end{botherref}
\endbibitem

\bibitem{loche2022surface}
\begin{barticle}
\bauthor{\bsnm{Loche}, \binits{M.}},
\bauthor{\bsnm{Scaringi}, \binits{G.}},
\bauthor{\bsnm{Yunus}, \binits{A.P.}},
\bauthor{\bsnm{Catani}, \binits{F.}},
\bauthor{\bsnm{Tanya{\c{s}}}, \binits{H.}},
\bauthor{\bsnm{Frodella}, \binits{W.}},
\bauthor{\bsnm{Fan}, \binits{X.}},
\bauthor{\bsnm{Lombardo}, \binits{L.}}:
\batitle{Surface temperature controls the pattern of post-earthquake landslide
  activity}.
\bjtitle{Scientific reports}
\bvolume{12}(\bissue{1}),
\bfpage{1}--\blpage{11}
(\byear{2022})
\end{barticle}
\endbibitem

\bibitem{hueckel2009explaining}
\begin{barticle}
\bauthor{\bsnm{Hueckel}, \binits{T.}},
\bauthor{\bsnm{Fran{\c{c}}ois}, \binits{B.}},
\bauthor{\bsnm{Laloui}, \binits{L.}}:
\batitle{Explaining thermal failure in saturated clays}.
\bjtitle{G{\'e}otechnique}
\bvolume{59}(\bissue{3}),
\bfpage{197}--\blpage{212}
(\byear{2009})
\end{barticle}
\endbibitem

\bibitem{Reichenbach2018}
\begin{barticle}
\bauthor{\bsnm{Reichenbach}, \binits{P.}},
\bauthor{\bsnm{Rossi}, \binits{M.}},
\bauthor{\bsnm{Malamud}, \binits{B.D.}},
\bauthor{\bsnm{Mihir}, \binits{M.}},
\bauthor{\bsnm{Guzzetti}, \binits{F.}}:
\batitle{A review of statistically--based landslide susceptibility models}.
\bjtitle{Earth--Science Reviews}
\bvolume{180},
\bfpage{60}--\blpage{91}
(\byear{2018}).
\doiurl{10.1016/j.earscirev.2018.03.001}
\end{barticle}
\endbibitem

\bibitem{shibasaki2016experimental}
\begin{barticle}
\bauthor{\bsnm{Shibasaki}, \binits{T.}},
\bauthor{\bsnm{Matsuura}, \binits{S.}},
\bauthor{\bsnm{Okamoto}, \binits{T.}}:
\batitle{Experimental evidence for shallow, slow-moving landslides activated by
  a decrease in ground temperature}.
\bjtitle{Geophysical Research Letters}
\bvolume{43}(\bissue{13}),
\bfpage{6975}--\blpage{6984}
(\byear{2016})
\end{barticle}
\endbibitem

\bibitem{shibasaki2017temperature}
\begin{barticle}
\bauthor{\bsnm{Shibasaki}, \binits{T.}},
\bauthor{\bsnm{Matsuura}, \binits{S.}},
\bauthor{\bsnm{Hasegawa}, \binits{Y.}}:
\batitle{Temperature-dependent residual shear strength characteristics of
  smectite-bearing landslide soils}.
\bjtitle{Journal of Geophysical Research: Solid Earth}
\bvolume{122}(\bissue{2}),
\bfpage{1449}--\blpage{1469}
(\byear{2017})
\end{barticle}
\endbibitem

\bibitem{mavsin2017coupled}
\begin{barticle}
\bauthor{\bsnm{Ma{\v{s}}{\'\i}n}, \binits{D.}}:
\batitle{Coupled thermohydromechanical double-structure model for expansive
  soils}.
\bjtitle{Journal of Engineering Mechanics}
\bvolume{143}(\bissue{9}),
\bfpage{04017067}
(\byear{2017})
\end{barticle}
\endbibitem

\bibitem{gariano2016landslides}
\begin{barticle}
\bauthor{\bsnm{Gariano}, \binits{S.L.}},
\bauthor{\bsnm{Guzzetti}, \binits{F.}}:
\batitle{Landslides in a changing climate}.
\bjtitle{Earth-Science Reviews}
\bvolume{162},
\bfpage{227}--\blpage{252}
(\byear{2016})
\end{barticle}
\endbibitem

\bibitem{elia2017numerical}
\begin{barticle}
\bauthor{\bsnm{Elia}, \binits{G.}},
\bauthor{\bsnm{Cotecchia}, \binits{F.}},
\bauthor{\bsnm{Pedone}, \binits{G.}},
\bauthor{\bsnm{Vaunat}, \binits{J.}},
\bauthor{\bsnm{Vardon}, \binits{P.J.}},
\bauthor{\bsnm{Pereira}, \binits{C.}},
\bauthor{\bsnm{Springman}, \binits{S.M.}},
\bauthor{\bsnm{Rouainia}, \binits{M.}},
\bauthor{\bsnm{Van~Esch}, \binits{J.}},
\bauthor{\bsnm{Koda}, \binits{E.}}, \betal:
\batitle{Numerical modelling of slope--vegetation--atmosphere interaction: an
  overview}.
\bjtitle{Quarterly Journal of Engineering Geology and Hydrogeology}
\bvolume{50}(\bissue{3}),
\bfpage{249}--\blpage{270}
(\byear{2017})
\end{barticle}
\endbibitem

\bibitem{hengl2017soilgrids250m}
\begin{barticle}
\bauthor{\bsnm{Hengl}, \binits{T.}},
\bauthor{\bparticle{de} \bsnm{Jesus}, \binits{J.M.}},
\bauthor{\bsnm{Heuvelink}, \binits{G.B.}},
\bauthor{\bsnm{Gonzalez}, \binits{M.R.}},
\bauthor{\bsnm{Kilibarda}, \binits{M.}},
\bauthor{\bsnm{Blagoti{\'c}}, \binits{A.}},
\bauthor{\bsnm{Shangguan}, \binits{W.}},
\bauthor{\bsnm{Wright}, \binits{M.N.}},
\bauthor{\bsnm{Geng}, \binits{X.}},
\bauthor{\bsnm{Bauer--Marschallinger}, \binits{B.}}, \betal:
\batitle{{SoilGrids250m: Global gridded soil information based on machine
  learning}}.
\bjtitle{PLoS one}
\bvolume{12}(\bissue{2}),
\bfpage{0169748}
(\byear{2017})
\end{barticle}
\endbibitem

\bibitem{kurylyk2015}
\begin{barticle}
\bauthor{\bsnm{Kurylyk}, \binits{B.}},
\bauthor{\bsnm{MacQuarrie}, \binits{K.}},
\bauthor{\bsnm{Caissie}, \binits{D.}},
\bauthor{\bsnm{McKenzie}, \binits{J.}}:
\batitle{Shallow groundwater thermal sensitivity to climate change and land
  cover disturbances: derivation of analytical expressions and implications for
  stream temperature modeling}.
\bjtitle{Hydrology and Earth System Sciences}
\bvolume{19},
\bfpage{2469}--\blpage{2489}
(\byear{2015})
\end{barticle}
\endbibitem

\end{thebibliography}



\end{document}